# Development of Machine learning algorithms to identify the Cobb angle in adolescents with idiopathic scoliosis based on lumbosacral joint efforts during gait (Case study)

Bahare Samadi [a, b], Maxime Raison [a, b], Philippe Mahaudens [c,d], Christine Detrembleur [d], Sofiane Achiche [a]

[a]Department of Mechanical Engineering Polytechnique Montréal, Canada; [b] Technopole in Pediatric Rehabilitation Engineering, Sainte-Justine UHC, Montreal, Canada; [c] Cliniques universitaires Saint-Luc, Service d'Orthopédie et de Traumatologie de l'Appareil Locomoteur, Brussels, Belgium ; [d] Université catholique de Louvain, Secteur des Sciences de la Santé, Institut de Recherche Expérimentale et Clinique, *Neuro Musculo Skeletal Lab (NMSK),* Brussels, Belgium

**Abstract-<u>Objectives:</u>** To quantify the magnitude of spinal deformity in adolescent idiopathic scoliosis (AIS), the Cobb angle is measured on X-ray images of the spine. Continuous exposure to X-ray radiation to follow-up the progression of scoliosis may lead to negative side effects on patients. Furthermore, manual measurement of the Cobb angle could lead to up to 10° or more of a difference due to intra/inter observer variation. Therefore, the objective of this study is to identify the Cobb angle by developing an automated radiation-free model, using Machine learning algorithms. **<u>Methods:</u>** Thirty participants with lumbar/thoracolumbar AIS (15° < Cobb angle < 66°) performed gait cycles. The lumbosacral (L5-S1) joint efforts during six gait cycles of participants were used as features to feed training algorithms. Various regression algorithms were implemented and run. **<u>Results:</u>** The decision tree regression algorithm achieved the best result with the mean absolute error equal to 4.6° of averaged 10-fold cross-validation. **<u>Conclusions:</u>** This study shows that the lumbosacral joint efforts during gait as radiation-free data are capable to identify the Cobb angle by using Machine learning algorithms. The proposed model can be considered as an alternative, radiation-free method to X-ray radiography to assist clinicians in following-up the progression of AIS.

## 1. Introduction

Adolescent idiopathic scoliosis (AIS), as the most common form of scoliosis, affects up to 4% of adolescents between ages 10 to 16 years old. It is defined by a deformation of the spine and trunk greater than 10 degrees [1]. The etiology is not clear; but various theories such as biomechanical, neuromuscular, genetic, and environmental origins exist [2]. The preliminary diagnosis is performed by screening with Adam's forward bend test [3] and a scoliometer [4]. The ultimate diagnosis and assessment is made by measuring the Cobb angle [5] as a gold standard. Cobb angle is the degree of spine curvature on the coronal plane, measured by using spinal radiography in a standing position. The accuracy of the measurement of the Cobb angle is an important basis for selecting therapeutic methods and following the progression of scoliosis. However, it is required to expose patients to sequential radiographic images i.e. at least every 6 months during their growth stage to monitor the progression of spinal deformity.

To define the Cobb angle, two main challenges exist; on the one hand, accurate Cobb angle measurement, utilizing the information of X-rays, requires considerable time and effort, along with associated problems such as inter/intra observer variations [6]. On the other hand, convolutional exposure to X-ray radiations in growing ages represents a real risk of an increase in organ carcinogenic and can cause tissue damage [7–10].

The difficulty in visualizing the vertebrae or poor selection of terminal vertebrae or variations in goniometers could decrease the measurement accuracy [11,12]. These increased the interest of using automated measurements methods such as Machine learning models and radiation-free data such as kinematics parameters instead of manual measurement and using radiographic data.

According to the state of the art methods, identifying the Cobb angle on X-ray or topography images using computer-assisted methods and Machine learning algorithms could provide an acceptable accuracy (mean absolute error (MAE) ranged between 1.9° and 7.8°) [11,13–17] compared to the manual measurement by specialized clinicians.

Consequently, it would be more efficient to develop an accurate radiation-free method without the need for topography and X-ray images. Seo *et al.* [18] predicted the Cobb angle with an MAE of 5° by applying multiple regression analysis on posture parameters in the standing position. They trained their model over 85% of the dataset (27 individuals) and tested it on 15% (5 individuals).

Gait analysis studies show that spinal deformity affects spino-pelvic mobility and can accordingly modify human locomotion and walking patterns [19]. Therefore, it is interesting to study the relationship between spinal deformity and gait pattern to identify the Cobb angle. Mahaudens et al. reported that scoliosis modifies the gait pattern and reduces the pelvic frontal motion[20]. Severe spinal deformity in AIS causes higher mobility resulting in higher joint efforts in the lumbosacral (L5-S1) joint than of healthy adolescents during gait [21]. Guilbert *et al.* [21] and Samadi *et al.*, [22,23], presented the differences of gait patterns among adolescents with AIS and typically developed adolescents by comparing the intervertebral efforts along the spine. All these studies attest to the major impact of scoliosis on mobility.

Accordingly, the aim of this study is to propose a radiation-free method, to identify the Cobb angle in AIS based on the lumbosacral joint efforts during gait, by using Machine learning algorithms.

## 2. Materials and methods

### 2.1 Dataset

Thirty AIS with main thoracolumbar/lumbar spine curvature (Lenke 5-6 [24]) with a Cobb angle between 15 and 66 degrees, without any history of spine surgery, participated in a previous study [20] were included in this research, as shown in Table 1. This longitudinal dataset is a collection that was started in 2009.

Table 1 Specification of the participants

| Cobb angle, degree (mean ± SD) | Age, years (mean ± SD) | Weight, kg (mean ± SD) | Height, cm (mean ± SD) |
|---|---|---|---|
| 35° ± 13° (Range: 15-66°) | 15 ± 2 | 50 ±9 | 163 ±9 |

This study takes ground on a previous study [19], which calculated lumbosacral joint efforts utilizing inverse dynamics. The features to train the Machine learning models have been selected based on the lumbosacral joint efforts, using reflective markers attached to specific anatomical joints on the skin of the participants performing gait cycles. A motion capture system composed of optokinetic sensors measured the kinematics of 26 anatomical landmarks. Ground reaction forces applied to each foot were obtained with an equipped treadmill with force sensors. The intervertebral forces and torques in the lumbosacral joint were calculated using a precise 3D inverse dynamic model of the human body, through ROBOTRAN software [19,25] as shown in Figure 1.

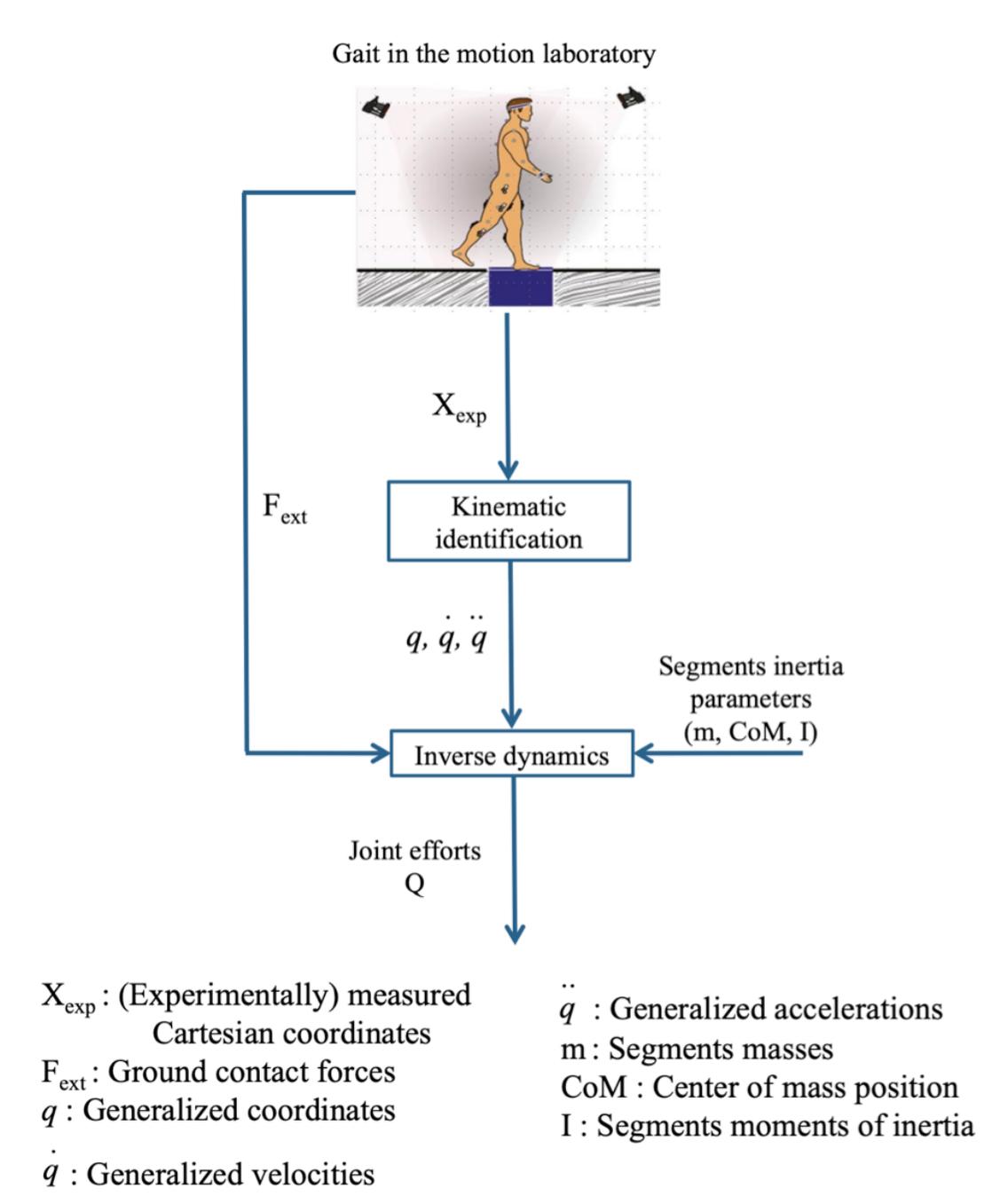

Figure 1 Process of quantifying the intervertebral joint efforts by inverse dynamics

## 2.2 Feature selection

According to the studies of the impact of the spinal deformity on gait patterns and intervertebral efforts, presented in [19–23] severe lumbar/thoracolumbar scoliosis presents higher mobility on the lumbosacral joint and therefore is led to different joint efforts in the lumbosacral joint, compared to the non-scoliosis spine. Furthermore, a single lumbar curve causes asymmetry in trunk motion in the coronal plane. As a result, the produced efforts during gait at the lumbosacral joint could be considered as significant parameters to define the scoliosis severity and to identify the Cobb angle.

Among the 3D intervertebral efforts, anteroposterior (AP) (the axis along with the walking direction) torques and mediolateral (ML) (the axis perpendicular to the walking direction) forces and torques are the ones which are affected the most by spinal deformity [21,23], therefore the features have been chosen based on these mentioned efforts.

We chose the optimal statistical combinations [26] from six gait cycles performed by each patient, shown in Table 2 as features to train our model.

Table 1 Features

| No. | Features | | |
|---|---|---|---|
| | **Mediolateral force** | **Anteroposterior torque** | **Mediolateral torque** |
| 1 | Maximum (\|Maximum-mean\|, \|Minimum-mean\|) | | |
| 2 | Maximum (\|Variance-mean\|, \|Variance+mean\|) | | |
| 3 | Maximum (\|Max\|, \|Min\|) | | |
| 4 | \|mean\| | | |
| 5 | Standard deviation | | |
| 6 | \|Max-Min\| | | |

## 2.3 Sampling process

To avoid any of the features dominating the model and learning algorithm as well as enabling the learning process to learn equally from all features, StandardScaler [27] was applied to the features. StandardScaler aligns the features in the same range of variance. The features were calculated over 360 steps i.e. 12 steps (6 complete successive gait cycles) for each participant. To account for the randomness of the splitting of the data and to evaluate the performance of each model on unseen data, k-fold cross-validation with 10 folds was applied on the algorithm [28]. The advantage of implementing cross-validation to train the model on a limited number of samples is that, it allows to generally estimate the performance of the model on data not used to train the model [28]. K-fold cross validation splits the data to k groups and each time it holds one group for the test, and it trains the model on the rest of the groups. The overall performance of the model is calculated based on the average of all groups over each time of the training process. K=10 has been shown empirically to yield to a model skill estimation with low bias and modest variance [28,29].

## 2.4 Model selection

Regression models, listed in Table 3, were developed to figure out the best model to identify the Cobb angle, feeding 18 features to the learning algorithms. To improve the accuracy of each model, Grid-search hyperparameter tuning was used. Hyperparameters are specified parameters to control the behaviour of Machine learning algorithms by tuning and Grid Search chooses the hyperparameters by going through all possible combinations of hyperparameters to obtain better prediction [30]. Table 3 presents the developed models with related parameters. The undefined parameters are the default parameters by scikit-learn software, which is a free Machine learning library in Python [31]. The model with the smallest MAE, which is the average MAE over the 10 folds, is the model with the best performance to identify the Cobb angle.

Table 02 Parameters used in each algorithm

| No. | Regression Algorithm | Parameters |
|---|---|---|
| 1 | K-nearest neighbours [32] | Number neighbours=3 (Number of the nearest neighbours in the dataset) |
| 2 | SVM with linear Kernel [33] | C = 100 (Regularization parameter, to avoid overfitting) |
| 3 | Random forests [34] | criterion=mae (The function to measure the quality of a split. mae : mean absolute error) max_depth=15 (The maximum depth of the tree) max_features=log2 (The number of features to consider when looking for the |

| | | best split. If "log2" is chosen, it means that it is equal to log2 of number of features) |
|---|---|---|
| 4 | Linear regression [35] | |
| 5 | Linear ridge regression [36] | alpha=0.1 (To define the regularization strength) |
| 6 | Lasso regression [37] | alpha=0.1 max_iter=10000 (The maximum number of iteration) selection=random (It accelerates the convergence) tol=0.0001 The tolerance for the optimization |
| 7 | Gaussian process [38] | kernel = DotProduct() + |

| | | WhiteKernel() |
| --- | --- | --- |
| | | (To specify the covariance function) |
| 8 | Multilayer Perceptron [37] | activation= logistic |
| | | (The function to activate the hidden layer, |
| | | The logistic sigmoid function: |
| | | $$f(x) = \frac{1}{1 + exp\,(-x)}$$ |
| | | hidden_layer_sizes=(60) |
| | | (Number of hidden neurons) |
| | | learning_rate=constant |
| | | (updates the weight) |
| | | learning_rate_init=0.003 |
| | | (controls the step-size in updating the weights) |

| | | | max_iter=1000 (number of epochs when the solver is "adam") n_iter_no_change=10 (number of iterations with no improvement to wait before termination training procedure) solver=adam (Optimizes the weight, "adam": stochastic gradient-based optimizer) |
|---|---|---|---|
| 9 | | AdaBoost [39] | n_estimators=250 (The maximum number of estimators at which boosting is terminated) learning_rate=1.1 (Shrinks the contribution of each |

| | | regressor) |
|---|---|---|
| | | loss=linear
(To update the weights after each boosting iteration) |
| 10 | Decision Tree [40] | max_depth = 4
The maximum depth of the tree

Criterion = "mse"
(The function to measure the quality of a split.
"mse" for the mean squared error, which is equal to variance reduction as feature selection criterion and minimizes the L2 loss using the mean of each terminal node) |
| 11 | Bootstrap aggregating (Bagging) [41] | base_estimator=Decision Tree
(The base estimator to fit on random subsets of the dataset) |

| | | |
|---|---|---|
| | | n_estimators=20

(The number of base estimators in the ensemble) |
| 12 | Gradient Boosting [42] | alpha=0.85

criterion=mae

learning_rate=1

learning rate shrinks the contribution of each

loss=huber

(Huber loss is a combination of both Mean Square Error and Mean Absolute Error)

max_depth=3

(maximum depth of the individual regression estimators |

| | | that limits the number of nodes in the tree) |
|---|---|---|

### 2.4.1 Decision Tree regression algorithm

As presented in Table 4, the decision tree model, presented the best result, therefore in this section, we go through the details of this model.

The decision tree regression model has an empirical tree structure that breaks down the dataset into smaller subsets by applying a series of simple rules learned from features, and at the same time, a related decision tree is gradationally grown. The gained information is then used to predict the target through a repetitive process of dividing [40,43]. To maximize the information gain (IG) at each split, an objective function is defined to optimize via the learning algorithm. The IG of a random variable $X$ obtained from an observation of a random variable $AA$ taking value $AA = a$ is defined in equation 1 [43]:

$$IG_{X,A}(X, a) = D_{KL}\big(P_X(x|a) \| P_X(x \vee I)\big) \qquad \text{Equation 1}$$

where,

$D_{KL}$ : Kullback–Leibler divergence which is a directed divergence between two distributions [44]

$P_X(x \vee I)$ : Prior distribution

$P_X(x|a)$ : Posterior distribution

## 3. RESULTS

### 3.1 Performance of regression models

The MAE (Equation 2) of each method which is the average MAE of 10 folds cross-validation between the measured Cobb angle by the clinicians using the radiography images and the ones predicted by each algorithm are shown in Table 4. The lowest MAE of 4.6° was obtained by implementing a decision tree model.

$$MAE = \frac{\sum_{i=1}^{n} y_i - x_i}{n} \qquad \text{Equation 2}$$

where,

$y_i$ : Prediction

$x_i$ : True value

$n$ : Total number of data points

Table 3   Mean absolute error (MAE) of Cobb angle, average of 10-fold cross validation for each model

*Algorithm with the best result

| No. | Regression Algorithm | Cross-Validation MAE (SD) degrees |
|---|---|---|
| 1 | K-nearest neighbours | 9.9 (5.1) |
| 2 | SVM with linear Kernel | 10.6 (3.6) |
| 3 | Random forests | 9.0 (5.3) |
| 4 | Linear regression | 11.9 (7.5) |
| 5 | Linear ridge regression | 10.5 (3.5) |
| 6 | Lasso regression | 9.4 (3.5) |
| 7 | Gaussian process | 8.9 (4.9) |
| 8 | Multilayer perceptron | 8.8 (2.4) |
| 9 | AdaBoost | 6.9 (3.9) |
| 10 | ***Decision Tree** | **4.6 (3.5)** |
| 11 | Bootstrap aggregating (Bagging) | 7.8 (5.3) |
| 12 | Gradient Boosting | 8.6 (5.6) |

The time of execution for all of the tested algorithms with the cross-validation and only the decision tree algorithm is 30.20s and 0.02s, respectively.

## 4. DISCUSSION

For the first time, we developed a radiation-free machine learning tool to identify the Cobb angle of AIS with lumbar/thoracolumbar spine curvature with an MAE of 4.6°. Furthermore, to our knowledge, this is the first study comparing regression models in this field. ML force and torque and AP torque in the lumbosacral joint as the most mobile part of the spine in lumbar and thoracolumbar scoliosis and the most affected intervertebral efforts by the spinal deformity [20,21,23], were fed to training algorithms. The process of Cobb angle identification was performed in a dynamic situation, i.e. walking which is a crucial activity of daily life.

Observing the value of obtained MAE from different models and compared to the variations in manual measurements which is up to 10° shows that as a preliminary study, the model has the potential of being used as an alternative to X-ray images. It is worth noting that using a systematic method could eliminate the variations in the manually Cobb angle measurement. For instance, radiography or body images have been used to feed Machine learning models to identify the Cobb angle [11,17,18,45,46], however it still kept the need for radiation exposure . Cumulative radiation exposure, i.e. 10 to 25 spinal X-rays during growth ages, increases the risk of cancer [47]. As reported by Simony *et al.* (2016) [48], the overall rate of cancer on 215 AIS patients was five times higher than the aged matched population and endometrial and breast cancer was most frequent. The dose of radiation applied to the patients in this mentioned study was comparable to modern equipment [48].

Observing the performance of applied models show that the value of the MAE in linear models is higher than the other models due to the fact that the data set does not have a fully linear relationship. The models which are the combination of other regression models i.e. AdaBoost (Adaptive Boosting) and Bootstrap aggregating (Bagging) have provided more accurate predictions as the ensemble model intends to improve the accuracy by reducing the variance. The decision tree provided the lowest value for MAE compared to the others. It can be explained due to its ability in feature selection automatically. In decision tree models, the accuracy of the model is not affected by the presence of multicollinearity (the presence of features that depend on each other) of the features [49].

In spite of interest in developing automated radiation-free methods as an alternative to X-ray, there are still very few studies in the field. Seo *et al.* [12] proposed a multiple regression analysis model using 21 static posture parameters to predict the Cobb angle with an MAE of 5° tested on only 5 patients without performing any cross-validation, however it is also known that spinal deformity affects mobility and locomotion. Cho *et al.* [45] reported 85.7% accuracy to classify the severity of scoliosis to less and more than 25° of the Cobb angle by applying the support vector machine method on 72 kinematic parameters during gait. Our proposed preliminary method as an automated radiation-free method, succeeded in identifying the Cobb angle with an MAE lower than similar studies during gait as a main daily activity.

This method would assist the clinicians to evaluate the progress of spinal deformity and improve the efficiency of treatments while reducing radiation exposure. As shown in Figure 2, manual measurement of the Cobb angle based on the radiography images of the spine may decrease the measurement accuracy up to 10° due to the difficulty in visualizing the vertebrae, poor selection of terminal vertebrae and variations in goniometers. For instance, as reported in [50–52] an error of 5°-10° may occur due to intra-observer and can exceed due to inter-observer variations, measured by the clinicians. Furthermore, radiographic acquisition and measurement methods could also lead to 2° to 7° error [53]. Therefore, an automatic method based on the gait pattern with an MAE of 4.6° could be considered as a reliable model with the potential of improving and implementing in determination of treatments strategies. However, the validation of the model on a larger dataset is required.

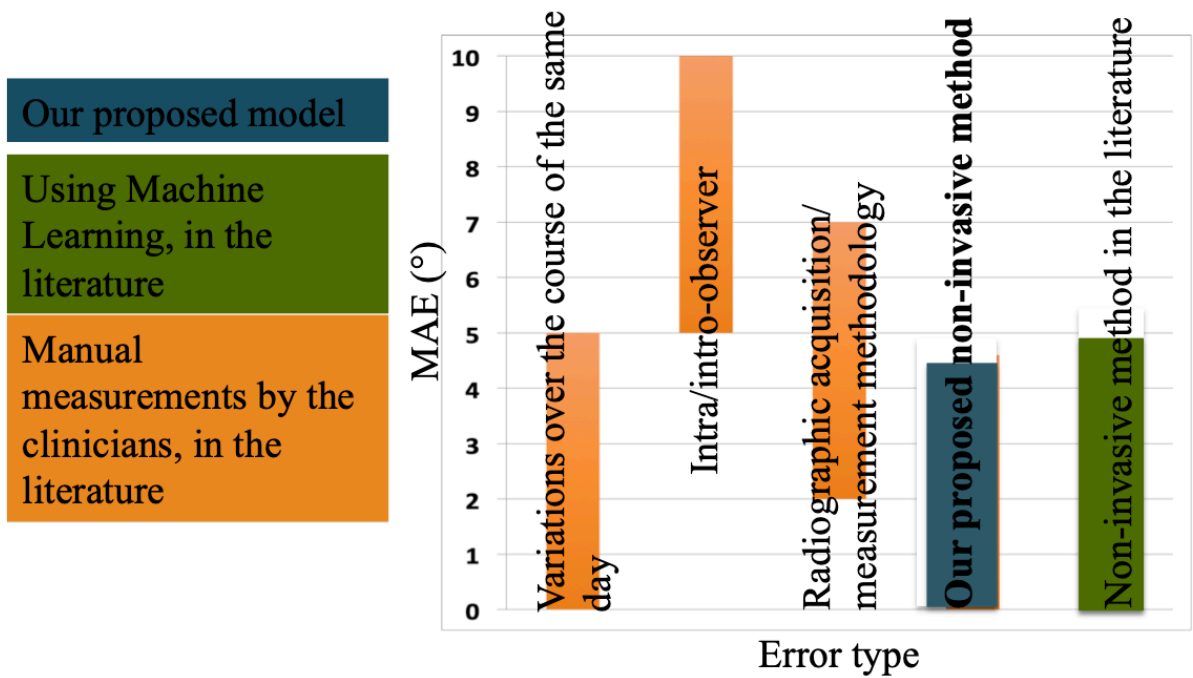

Figure 2 Comparison between the Cobb angle measurement errors and the error of the proposed model [18,50–53]

The dataset included in the present work were obtained from adolescents with AIS with thoracolumbar/lumbar curves. Therefore, future studies should be performed to test the method on other type of AIS i.e. the curvature located on the thoracic part of the spine.

[25] Robotran - Home n.d. http://www.robotran.be/ (accessed December 3, 2018).

[26] Guo C, Lu M, Chen J. An evaluation of time series summary statistics as features for clinical prediction tasks. BMC Medical Informatics and Decision Making 2020;20:48. https://doi.org/10.1186/s12911-020-1063-x.

[27] D.k. T, B.g P, Xiong F. Auto-detection of epileptic seizure events using deep neural network with different feature scaling techniques. Pattern Recognition Letters 2019;128:544–50. https://doi.org/10.1016/j.patrec.2019.10.029.

[28] Browne MW. Cross-Validation Methods. Journal of Mathematical Psychology 2000;44:108–32. https://doi.org/10.1006/jmps.1999.1279.

[29] Berrar D. Cross-Validation, 2018. https://doi.org/10.1016/B978-0-12-809633-8.20349-X.

[30] Bardenet R, Brendel M, Kégl B, Sebag M. Collaborative hyperparameter tuning. Proceedings of the 30th International Conference on International Conference on Machine Learning - Volume 28, Atlanta, GA, USA: JMLR.org; 2013, p. II-199-II–207.

[31] Pedregosa F, Varoquaux G, Gramfort A, Michel V, Thirion B, Grisel O, et al. Scikit-learn: Machine Learning in Python. Journal of Machine Learning Research 2011;12:2825–30.

[32] sklearn.neighbors.KNeighborsRegressor — scikit-learn 0.23.2 documentation n.d. https://scikit-learn.org/stable/modules/generated/sklearn.neighbors.KNeighborsRegressor.html (accessed August 29, 2020).

[33] sklearn.svm.SVR — scikit-learn 0.23.2 documentation n.d. https://scikit-learn.org/stable/modules/generated/sklearn.svm.SVR.html (accessed August 29, 2020).

[34] 3.2.4.3.2. sklearn.ensemble.RandomForestRegressor — scikit-learn 0.23.2 documentation n.d. https://scikit-learn.org/stable/modules/generated/sklearn.ensemble.RandomForestRegressor.html (accessed August 29, 2020).

[35] sklearn.linear_model.LinearRegression — scikit-learn 0.23.2 documentation n.d. https://scikit-learn.org/stable/modules/generated/sklearn.linear_model.LinearRegression.html (accessed September 29, 2020).

[36] sklearn.linear_model.Ridge — scikit-learn 0.23.2 documentation n.d. https://scikit-learn.org/stable/modules/generated/sklearn.linear_model.Ridge.html (accessed August 31, 2020).

[37] sklearn.linear_model.Lasso — scikit-learn 0.23.2 documentation n.d. https://scikit-learn.org/stable/modules/generated/sklearn.linear_model.Lasso.html (accessed September 29, 2020).

[38] sklearn.gaussian_process.GaussianProcessRegressor — scikit-learn 0.23.2 documentation n.d. https://scikit-learn.org/stable/modules/generated/sklearn.gaussian_process.GaussianProcessRegressor.html (accessed August 29, 2020).
28